\documentclass[aps,prd,10pt,twocolumn,superscriptaddress,nofootinbib,floatfix]{revtex4-2}
\usepackage{graphicx}  
\usepackage{amsmath,amssymb,amsfonts}
\usepackage{bm}        
\usepackage{physics}   
\usepackage{mathrsfs}  
\usepackage{booktabs}  
\usepackage{longtable}
\usepackage{array}
\usepackage{float}
\usepackage{enumitem}
\usepackage{hyperref}  
\hypersetup{
    colorlinks=true,
    linkcolor=black,
    citecolor=black,
    urlcolor=black
}

\begin{document}

\title{The final fate of anisotropic-dissipative gravitational collapse}

\author{Kanabar Jay }
\affiliation{MG Science Institute, Gujarat University, Ahmedabad 380009, India}
\affiliation{International Centre for Space and Cosmology, Ahmedabad University, Ahmedabad 380009, India}
\email{kanabarjaymgsi@gmail.com}

\date{\today}


\begin{abstract}
The final fate of a collapsing star depends not only on how much matter it contains but also on how that matter resists gravity in different directions. In this work, we investigate the final fate of highly magnetized radiation-dominated spherically symmetric dissipative stellar configurations. We study the dynamics of collapse by introducing a dimensionless geometric factor $f(\theta, \phi)$ defined by the angular dependence of the radiative opacity. Using the field equations, we derive direction-sensitive threshold conditions that determine whether collapse initiates, halts, or reverses. The resulting inequalities unify black hole formation, bounce behavior, and delayed trapping of geodesics into a single geometrically controlled framework. This theoretical analysis would help analyze the collapse through pre-computed opacity tables for different magnetic field, temperature, and density profiles, along with other data available for such considered profiles.
\end{abstract}

\maketitle

\section{Introduction}
The final fate of the gravitational collapse of a massive body is one of the most fundamental questions in gravitation theory and relativistic astrophysics. What determines whether a collapsing star forms a black hole, or if it halts and turns into a bounce, or ends in a visible singularity? The classical gravitational collapse model given by Oppenheimer and Snyder \cite{oppenheimer1939} in 1939 and independently by Dutt \cite{datt1938} (OSD) in 1938, which describes the collapse of a pressure-less homogeneous dust cloud, showed that sufficiently massive matter configurations inevitably evolve towards the formation of a spacetime singularity covered by an event horizon. This led to the formulation of singularity theorems by Penrose \cite{penrose1965} which demonstrated the inevitability of singularities in gravitational collapse under generic initial conditions. In their original formulations, the singularity theorems do not explicitly address the nature or visibility of the singularity. This introduces a critical ambiguity concerning the causal structure of spacetime: whether the singularities predicted are shielded by event horizons, as in black hole solutions, or whether they can be exposed to causal contact with local or global observers, leading to the formation of naked singularities. This ambiguity motivated the formulation of the cosmic censorship conjecture (CCC) by Penrose \cite{penrose1969}, which posits that nature forbids the formation of naked singularities. 

Despite decades of effort, CCC has only been studied for a few models, and a variety of models now exist where naked singularities form from generic initial conditions summarized in \cite{joshi2007,joshi20007}, suggesting that singularities might be experimentally accessible. In parallel, another line of investigation has explored mechanisms by which collapse may halt or reverse before reaching a singularity. One such class involves quantum gravitational bounces ~\cite{Malafarina2017}, and demonstrated in pressure-driven loop quantum gravity models~\cite{Cafaro2025}. Another class invokes exotic matter or negative pressures, where models show that trapped surface formation can be significantly delayed or avoided ~\cite{JoshiGoswami2007a, JoshiGoswami2007b}. These scenarios collectively point towards a richer and broader range of possible gravitational collapse outcomes beyond the classical black hole paradigm. 

One recurring theme in these diverse collapse scenarios is the critical role of pressure anisotropy and dissipation. The pioneering work of Herrera and Santos \cite{herrera1997,herrera2006} introduced anisotropic fluid models, demonstrating that even within spherical symmetry, anisotropy generates additional internal stresses that significantly modify both equilibrium and dynamical evolution. Extensions of this framework have incorporated viscosity, shear, and radial heat flux \cite{herrera2009,herrera2018fluids}, allowing for temporary halting or even reversal of collapse. These studies suggest that collapse is not governed solely by the total energy content or symmetry assumptions, but also by the internal directional dynamics of the matter field with regard to how pressure, radiation, and transport respond to gravitational curvature. 

Anisotropic models offer a more realistic representation of compact stellar interiors, where equilibrium is often disturbed by temperature gradients, magnetic fields, or radiation streaming. This opens the path to identifying concrete, physically motivated sources of anisotropy that can be systematically modeled and constrained. A particularly realistic and well-established source of pressure anisotropy arises in radiation-dominated, magnetized stellar environments, such as those found in proto-neutron stars, magnetars, and accretion-driven compact objects. In these settings, the radiative opacity, which quantifies the coupling between radiation and matter, becomes intrinsically direction-dependent due to the presence of strong magnetic fields, anisotropic scattering cross-sections, and steep temperature gradients. This angular variation in opacity leads to anisotropic momentum transfer within the stellar medium, which manifests macroscopically as pressure anisotropy in the fluid’s stress-energy tensor. This effect is well-documented in the context of stellar structure and evolution. To account for it, modern astrophysical modeling relies on pre-computed Rosseland and Planck mean opacity tables defined for thin and thick optical media respectively \cite{rogers1992,laming1993,barfield2002,potekhin2000a,potekhin2000b,mendoza2007,potekhin2014,mushtukov2015}, which incorporate detailed microphysics including electron scattering, free–free and bound–free transitions, magnetic polarization effects, and radiative transfer in various geometries. These tables provide opacity values for different temperatures, density profiles, and different magnetic field strengths, which are widely used in the modeling of neutron star crusts, magnetized white dwarfs, and high-energy stellar phenomena. 

In this work, we try to answer a fundamental and physically motivated question arising from our literature review: Can we determine the final fate of gravitational collapse, including black hole formation, bounce, or visible singularity, based purely on the angular structure of radiative opacity?
This can be further simplified to asking: Can we use the pre-computed opacity tables and stellar profiles to predict the end state of their gravitational collapse? In this work, we aim to answer these questions to probe the collapse dynamics through pre-computed opacity tables and similar stellar data available for such desired profiles. Using Einstein equations, we derive direction-sensitive threshold conditions that govern whether collapse initiates, accelerates, halts, or reverses. Importantly, we embed the angular dependence of the opacity function into the collapse dynamics, linking the theoretical framework to available stellar data. 

This work is structured as follows. In Section 2, we study the spacetime structure and dynamical equations of anisotropic collapse in the framework of general relativity. In Section 3, we define the geometric factor \( f(\theta, \phi) \) from angular-dependent opacity and express the anisotropic stress \( \Delta \) in terms of microphysical input. In Section 4, we analyze the full dynamical evolution equation and derive the directional threshold conditions that govern the onset of collapse, the halting of infall, and the bounce. These inequalities reveal how anisotropic radiation flow dictates the local gravitational behavior of the collapsing star. In Section 5, we interpret the physical consequences of these results, examining how the geometry of radiation transport can decide the final fate of the collapse. Finally, we discuss how our framework can be applied to real astrophysical systems by directly incorporating pre-computed opacity data from stellar evolution tables. In addressing the formation or avoidance of trapped surfaces, our analysis relies not on direct null geodesic integration but rather on dynamical conditions sourced from the stress-energy tensor — particularly those involving anisotropy and opacity-driven dissipation. We justify this approach in Appendix A, where we draw on established results demonstrating that sufficiently strong dissipation or radiative processes can delay or even prevent trapped surface formation. This physically motivated reasoning underpins our exploration of opacity-governed collapse. The approach developed here links the microphysical properties of radiation transport to the macroscopic dynamics of gravitational collapse in a covariant and observationally relevant way.

\section{Spacetime structure and dynamical equations of anisotropic collapse}
In this section we provide a review of the basic general relativistic framework required to study our desired collapsing configuration. We will construct our framework in a comoving coordinate frame.  We consider a spherically symmetric distribution of collapsing fluid, locally anisotropic, undergoing dissipation in the form of heat flow and free streaming radiation, bounded by a spherical surface \( \Sigma \).
For such system the energy–momentum tensor is given by, 
\begin{align}
T^{\alpha\beta} 
&= (\mu + P_\perp) V^\alpha V^\beta 
+ P_\perp g^{\alpha\beta} \nonumber \\
&\quad + (P_r - P_\perp) \chi^\alpha \chi^\beta  + q^\alpha V^\beta \nonumber \\
&\quad + V^\alpha q^\beta  + \epsilon\, l^\alpha l^\beta \ \ ,
\label{eqn1:Tmunu}
\end{align}
where $\mu$ is the energy density, $P_r$ the radial pressure, $P_\perp$ the tangential pressure, $\epsilon$ the radiation density, $V^\alpha$ is the four-velocity of the fluid, $q^\alpha$ is the heat flux, $\chi^\alpha$ is a unit four-vector along the radial direction, and $l^\alpha$ is a null four-vector. These quantities satisfy
\begin{align}
V^\alpha V_\alpha &= -1, &\quad V^\alpha q_\alpha &= 0, \nonumber \\
\chi^\alpha \chi_\alpha &= 1, &\quad \chi^\alpha V_\alpha &= 0, &\quad l^\alpha l_\alpha &= 0.
\label{eqn2:orthogonality}
\end{align}
We assume the interior metric to be comoving, shear-free for simplicity, and spherically symmetric. The line element of such metric is given as :
\begin{align}
ds^2 = & \quad-A^2(t,r)\,dt^2 \nonumber \\ 
& \quad+ B^2(t,r)\left(dr^2 + r^2 d\theta^2 + r^2 \sin^2\theta\, d\phi^2\right),
\label{eqn3:metric}
\end{align}
and hence,
\begin{align}
V^\alpha &= A^{-1} \delta^\alpha_0, \quad q^\alpha = q\, \delta^\alpha_1,\nonumber \\ 
 \quad l^\alpha &= A^{-1} \delta^\alpha_0 + B^{-1} \delta^\alpha_1, \quad \chi^\alpha = B^{-1} \delta^\alpha_1.
\label{eqn4:vectors}
\end{align}
Using Eqns. \eqref{eqn1:Tmunu},\eqref{eqn2:orthogonality},\eqref{eqn3:metric}, and \eqref{eqn4:vectors}, the Einstein field equations reduce to the following form:
\begin{widetext}
\begin{align}
8\pi T_{00} &= 8\pi(\mu + \epsilon) A^2 \nonumber\\
&= -\left(\frac{A}{B}\right)^2 \left[ 2\frac{B''}{B} - \left(\frac{B'}{B}\right)^2 + \frac{4}{r} \frac{B'}{B} \right] + 3\left(\frac{\dot{B}}{B}\right)^2,
\label{eq:Einstein00} \\
8\pi T_{01} &= -8\pi(qB + \epsilon)AB = -2\left[\frac{\dot{B}'}{B} - \frac{B'}{B} \frac{\dot{B}}{B} - \frac{A'}{A} \frac{\dot{B}}{B} \right],
\label{eq:Einstein01} \\
8\pi T_{11} &= 8\pi(P_r + \epsilon) B^2 = \left(\frac{B'}{B}\right)^2 + \frac{2}{r} \frac{B'}{B} + 2 \frac{A'}{A} \frac{B'}{B} + \frac{2}{r} \frac{A'}{A} \nonumber \\
&\quad - \left( \frac{B}{A} \right)^2 \left[ 2 \frac{\ddot{B}}{B} + \left( \frac{\dot{B}}{B} \right)^2 - 2 \frac{\dot{A}}{A} \frac{\dot{B}}{B} \right],
\label{eq:Einstein11} \\
8\pi T_{22} &= \frac{8\pi}{\sin^2 \theta} T_{33}^{-} = 8\pi r^2 P_\perp B^2 \nonumber\\
&= r^2 \left[ \frac{B''}{B} - \left( \frac{B'}{B} \right)^2 + \frac{1}{r} \frac{B'}{B} + \frac{A''}{A} + \frac{1}{r} \frac{A'}{A} \right] \nonumber \\
&\quad - r^2 \left( \frac{B}{A} \right)^2 \left[ 2\frac{\ddot{B}}{B} + \left( \frac{\dot{B}}{B} \right)^2 - 2 \frac{\dot{A}}{A} \frac{\dot{B}}{B} \right].
\label{eq:Einstein22}
\end{align}
\end{widetext}
The rate of expansion $\Theta = V^\alpha{}_{;\alpha}$ of the fluid sphere is given, from Eqns. \eqref{eqn3:metric} and \eqref{eqn4:vectors}, by
\begin{equation}
\Theta = \frac{3 \dot{B}}{AB},
\label{eq:expansion}
\end{equation}
and from Eqn. \eqref{eq:Einstein01}, we have
\begin{equation}
8\pi(qB + \epsilon)B = \frac{2}{3} \Theta'.
\label{eq:heatTheta}
\end{equation}

From Eqn. \eqref{eq:heatTheta}, if $q > 0$ and $\epsilon > 0$, then $\Theta' > 0$, meaning that if the system is collapsing ($\Theta < 0$), $q$ and $\epsilon$ reduce the rate of collapse toward the outer layers. If $q = 0$ and $\epsilon = 0$, then $\Theta' = 0$, implying homogeneous collapse. The mass function $m(t,r)$ is taken as given in Ref. \cite{cahill1970} and is obtained from the Riemann tensor component $R^2{}_{323}$ and, for the metric given in Eqn. \eqref{eqn3:metric}, is given by
\begin{equation}
m(t, r) = \frac{(rB)^3}{2} R^2{}_{323} = \frac{r^3}{2} \left( \frac{\dot{B}^2B}{A^2} - \frac{B'^2}{B} \right) - r^2 B'.
\label{eq:massfunction}
\end{equation}

\subsection{The exterior spacetime}
The exterior spacetime to $\Sigma$ of the collapsing body is described by the outgoing Vaidya spacetime, which models a radiating star and has the line element:
\begin{align}
ds^2_{+} = &-\left(1 - \frac{2m(v)}{\rho} \right) dv^2 \nonumber \\
           &- 2\, dv\, d\rho + \rho^2 \left( d\theta^2 + \sin^2 \theta\, d\phi^2 \right),
\label{eq:vaidya_metric}
\end{align}
where $m = m(v)$ is the total mass inside $\Sigma$, and is a function of the retarded time $v$. The surface $\Sigma$ is defined as $r = r_\Sigma = \text{constant}$ in the comoving coordinate system given in Eqn. \eqref{eqn3:metric}, and $\rho = \rho_\Sigma(v)$ in the non-comoving Vaidya coordinate system given by Eqn. \eqref{eq:vaidya_metric}. Matching the interior metric given by Eqn. \eqref{eqn3:metric} with the energy–momentum tensor given by Eqn. \eqref{eqn1:Tmunu} to the exterior metric \eqref{eq:vaidya_metric} using Darmois junction conditions yields the following matching conditions:
\begin{equation}
\left. P_r \right|_{\Sigma} = \left. q B \right|_{\Sigma},
\label{eq:junction1}
\end{equation}
\begin{equation}
\left. (qB + \epsilon) \right|_{\Sigma} = \left. \frac{1}{4\pi} \left( \frac{L}{\rho^2} \right) \right|_{\Sigma},
\label{eq:junction2}
\end{equation}
\begin{equation}
(rB)|_{\Sigma} = \rho_\Sigma,
\label{eq:junction3}
\end{equation}
\begin{equation}
\left. \left[ \frac{r^3}{2} \left( \frac{\dot{B}^2B}{A^2} - \frac{B'^2}{B} \right) - r^2 B' \right] \right|_{\Sigma} = m(v),
\label{eq:junction4}
\end{equation}
\begin{equation}
A_\Sigma\, dt = \left[ 1 - \frac{2m}{\rho} + 2 \frac{d\rho}{dv} \right]^{1/2}_{\Sigma} dv,
\label{eq:junction5}
\end{equation}
where $L$ is defined as the total luminosity of the collapsing sphere measured on its surface, given by
\begin{equation}
L = L_\infty \left( 1 - \frac{2m}{\rho} + 2\frac{d\rho}{dv} \right)^{-1},
\label{eq:luminosity_surface}
\end{equation}
and,
\begin{equation}
L_\infty = \frac{dm}{dv},
\label{eq:luminosity_infinity}
\end{equation}
is the luminosity measured by an observer at rest at infinity. Eqn. \eqref{eq:junction1} expresses the continuity of the radial flux of momentum across $\Sigma$, where only the heat flux $q$ appears. However, in the total radiation leaving $\Sigma$, Eqn. \eqref{eq:junction2} shows that the radiation $\epsilon$ contributes as well as $q$. It is important to observe that $\epsilon$ has the same null property associated with the exterior radiation it produces, while $q$ also contributes to the exterior radiation, and is not itself null. Eqn. \eqref{eq:junction3} enforces the continuity of the proper radius across $\Sigma$ as measured from the perimeter in both coordinate systems. Eqn. \eqref{eq:junction4} relates to the mass function given in Eqn. \eqref{eq:massfunction}, and Eqn. \eqref{eq:junction5} links the proper time across $\Sigma$ in both coordinate systems.

\subsection{Generalized dynamical equation and coupled dynamical–transport system}
We summarize the dynamical framework for analyzing relativistic gravitational collapse under dissipation approximation, based on the Misner–Sharp formalism and the causal transport theory of Müller–Israel–Stewart. For full derivations we refer to Herrera and Santos (2004) \cite{herrera2004dissipative}. The radial acceleration of a collapsing fluid element, measured by a comoving observer, is governed by the equation:
\begin{widetext}
\begin{equation}
(\mu + P_r + 2\epsilon)\, D_t U = -(\mu + P_r + 2\epsilon) \left[ \frac{m + 4\pi (P_r + \epsilon) R^3}{R^2} \right] 
- E^2 \left[ D_R(P_r + \epsilon) + \frac{2(\epsilon + P_r - P_\perp)}{R} \right] 
- E \left[ \frac{(5qB + 4\epsilon)U}{R} + B D_t q + D_t \epsilon \right],
\label{eq:dynamical_eq}
\end{equation}
\end{widetext}
where $U = r \ D_t B$ is the velocity of the collapsing fluid ($U < 0$ in the collapsing regime), $R = rB$ is the areal radius, and $D_t$ and $D_R$ denote derivatives with respect to proper time and proper radius, respectively. The quantity $E$ represents a generalized Lorentz-type energy factor, and $m(t,r)$ is the Misner–Sharp mass enclosed within radius $R$. This equation takes the familiar Newtonian-like structure:
\begin{equation}
\text{Force} = \text{Mass density} \times \text{Acceleration}.
\label{eq:newtonian_form}
\end{equation}
Following this structure, the first term on the right-hand side of Eqn.~\eqref{eq:dynamical_eq} represents the effective gravitational attraction acting on a fluid element. It includes not only the conventional Misner–Sharp mass $m$ but also relativistic contributions due to pressure and radiation, specifically $P_r$ and $\epsilon$. The second term corresponds to hydrodynamical forces: the radial pressure gradient $D_R(P_r + \epsilon)$ counteracts collapse, while the anisotropic contribution $2(P_r - P_\perp)/R$ may either assist or resist it, depending on the sign of the anisotropy. The third term encapsulates the dynamical role of dissipation. The term $(5qB + 4\epsilon)U/R$ is strictly positive in a collapsing regime ($U < 0$), indicating that both heat flux and null radiation reduce the net inward acceleration by extracting energy from the fluid interior. Notably, the heat flux $q$ contributes solely to this energy outflow and does not appear in gravitational or pressure terms, whereas the radiation density $\epsilon$ plays a dual role: it not only acts as a pressure source but also contributes to the active gravitational mass. The time derivatives $D_t q$ and $D_t \epsilon$ reflect changes in dissipative fluxes and can dynamically influence the collapse evolution. Altogether, Eqn.~\eqref{eq:dynamical_eq} captures the interplay between gravity, internal stresses, and irreversible processes in shaping the fate of a collapsing relativistic configuration.
In the static, non-dissipative limit (i.e., $U = q = \epsilon = 0$), Eqn.~\eqref{eq:dynamical_eq} reduces to the generalized Tolman–Oppenheimer–Volkoff (TOV) equation for anisotropic fluids:
\begin{equation}
D_R P_r + \frac{2(P_r - P_\perp)}{R} = -\frac{(\mu + P_r)}{R(R - 2m)} \left[ m + 4\pi P_r R^3 \right],
\label{eq:tov}
\end{equation}
which expresses equilibrium between the gravitational pull and the combined effect of radial and transverse pressure gradients. The second term on the left-hand side represents the effect of anisotropy: if $P_r > P_\perp$, it increases the effective pressure support against gravity; otherwise, it enhances collapse. When dissipation in the diffusion approximation is incorporated, the transport equation from Müller–Israel–Stewart theory couples with the dynamical equation and modifies the evolution. The resulting equation becomes:
\begin{widetext}
\begin{equation}
(\mu + P_r)(1 - \alpha)\, D_t U = F_{\text{grav}}(1 - \alpha) + F_{\text{hyd}} + \frac{E\kappa T'}{\tau B} + \frac{E q B}{\tau} + \frac{\kappa E T^2 q B}{2 A \tau} \left( \frac{d}{dt} \left( \frac{\tau}{\kappa T^2} \right) \right) - \frac{5 U E B q}{R},
\label{eq:dynamical_transport}
\end{equation}
\end{widetext}

where $\kappa$ is the thermal conductivity, $T$ the local temperature, $\tau$ the relaxation time for heat flux, and $A$ the lapse function from the interior metric. This equation makes clear that dissipation affects both inertia and gravity through the critical factor:
\begin{equation}
\alpha = \frac{\kappa T}{\tau(\mu + P_r)}.
\label{eq:alpha}
\end{equation}
As $\alpha \rightarrow 1$, the effective inertial mass density $(\mu + P_r)(1 - \alpha)$ tends to zero. Simultaneously, the effective gravitational term is reduced by the same factor. This situation marks a potential critical point, beyond which the collapsing fluid may bounce or rapidly decelerate, even under small perturbations. Physically, this reflects the loss of causal responsiveness of the fluid under extreme thermal conduction and short relaxation times, precisely the conditions expected in late-stage supernova cores or neutrino-trapped interiors. The gravitational force density appearing in Eqn.~\eqref{eq:dynamical_transport} is given explicitly by
\begin{equation}
F_{\text{grav}} = -(\mu + P_r) \left[ \frac{m + 4\pi P_r R^3}{R^2} \right],
\label{eq:Fgrav}
\end{equation}
which includes the general relativistic contribution of pressure to the gravitational mass. Meanwhile, the hydrodynamic force term is given by:
\begin{equation}
F_{\text{hyd}} = -E^2 \left[ D_R P_r + \frac{2(P_r - P_\perp)}{R} \right],
\label{eq:Fhyd}
\end{equation}
where again the second term highlights the role of pressure anisotropy. The direction and magnitude of this term can either resist or enhance the collapse, depending on the sign of $P_r - P_\perp$.

In summary, Eqns.~\eqref{eq:dynamical_eq}–\eqref{eq:Fhyd} encode the rich interplay between gravity, thermodynamics, pressure anisotropy, and dissipative processes in relativistic collapse. Dissipation not only alters the mass-energy content via heat and radiation outflows but also fundamentally modifies the inertial and gravitational response of the system via the $\alpha$ parameter. The system is particularly sensitive near $\alpha \to 1$, where even weak gradients or fluxes may trigger significant dynamical effects, including the possibility of a bounce.
\subsection{Physical conditions and collapse outcomes} 
The final outcome of gravitational collapse depends critically on the balance between local anisotropic stresses and the dynamical evolution of the collapsing matter. Specifically, the collapse may result in a black hole, a naked singularity, or a bounce, depending on the evolution of the radial velocity $U$ and its proper time derivative $D_t U$. For black hole formation, the collapse proceeds with $U < 0$ and accelerates with $D_t U < 0$. In contrast, a naked singularity forms when the collapse continues with $U < 0$ but decelerates with $D_t U > 0$, such that the formation of trapped surfaces is delayed beyond the singularity formation. A bounce scenario occurs when the collapse halts at $U = 0$ and an outward acceleration begins with $D_t U > 0$. In the table given below, we summarize these criteria by deriving the critical inequalities on pressure anisotropy following from some rearrangements of Eqn. (\ref{eq:dynamical_eq}), which would be the base of our analysis to embed the radiative opacity into the collapse dynamics. The conditions for deriving the singularity curve and the apparent horizon curve functions need to be studied separately for the naked singularity case, which is our desired outcome, since they are not explicit functions of anisotropy. However, in this work, we lay down the sufficient and necessary conditions for them. For full numerical simulations or data analysis, the energy density $\epsilon$ and $P_r$ can be replaced as a function of anisotropy as per the choice of the Equation of State. 
\begin{widetext}
\begin{table}[H] 
\centering
\renewcommand{\arraystretch}{1.6}
\begin{tabular}{|c|p{6.5cm}|p{7.5cm}|}
\hline
\textbf{Outcome} & \textbf{Physical Conditions} & \textbf{Critical Anisotropy Inequality on $\Delta$} \\
\hline

\textbf{Black Hole} & 
\begin{itemize}
    \item $U < 0$ (collapse proceeds)
    \item $D_t U < 0$ (collapse accelerates)
\end{itemize} 
& 
$\displaystyle \Delta < \frac{R}{2 E^2} \left[ 
\begin{aligned}
    & (\mu + P_r + 2\epsilon) \left( \frac{m + 4\pi (P_r + \epsilon) R^3}{R^2} \right) \\
    & + E^2 \left( D_R (P_r + \epsilon) + \frac{2\epsilon}{R} \right) \\
    & + E \left( 5B(q + \epsilon)\frac{U}{R} + B D_t q + D_t \epsilon \right)
\end{aligned}
\right]$ \\

\hline

\textbf{Naked Singularity} & 
\begin{itemize}
    \item $U < 0$ (collapse continues)
    \item $D_t U > 0$ (decelerating collapse)
    \item Trapping delayed beyond singularity
\end{itemize} 
& 
$\displaystyle \Delta > \frac{R}{2 E^2} \left[ 
\begin{aligned}
    & (\mu + P_r + 2\epsilon) \left( \frac{m + 4\pi (P_r + \epsilon) R^3}{R^2} \right) \\
    & + E^2 \left( D_R (P_r + \epsilon) + \frac{2\epsilon}{R} \right) \\
    & + E \left( 5B(q + \epsilon)\frac{U}{R} + B D_t q + D_t \epsilon \right)
\end{aligned}
\right]$ \\

\hline

\textbf{Bounce} & 
\begin{itemize}
    \item $U = 0$ (collapse halts)
    \item $D_t U > 0$ (outward acceleration begins)
\end{itemize} 
& 
$\displaystyle \Delta > \frac{R}{2 E^2} \left[ 
\begin{aligned}
    & (\mu + P_r + 2\epsilon) \left( \frac{m + 4\pi (P_r + \epsilon) R^3}{R^2} \right) \\
    & + E^2 \left( D_R (P_r + \epsilon) + \frac{2\epsilon}{R} \right) \\
    & + E \left( B D_t q + D_t \epsilon \right)
\end{aligned}
\right]$ \\
\hline
\end{tabular}
\caption{Physical conditions and critical anisotropy inequalities determining the gravitational collapse outcomes. Here, $\Delta = P_{\perp} - P_r$ denotes the local anisotropic stress, where $P_{\perp}$ is the tangential (perpendicular) pressure and $P_r$ is the radial (parallel) pressure within the collapsing matter distribution.}
\label{tab:anisotropy_conditions}
\end{table}
\end{widetext}

\section{\texorpdfstring{Geometric Factor \textit{f}($\theta$,$\phi$) and Anisotropic stress $\Delta$}
{Geometric Factor f(theta,phi) and Anisotropic stress Delta}}

We now derive the geometric factor to link the angular-dependent radiative opacity to collapse dynamics via spherical harmonics. The method to encode directional dependent opacity may change as one chooses other spacetimes which do not obey spherical symmetry.
 We define it as the ratio of the opacity in a general direction \( (\theta, \phi) \) to the parallel opacity \( \kappa_\parallel \):
\begin{equation}
f(\theta, \phi) = \frac{\kappa(\theta, \phi)}{\kappa_\parallel}
\label{eq:1}
\end{equation}
where opacity in any general direction \( (\theta, \phi) \) is given by:
\begin{equation}
    \kappa(\theta, \phi) = \sum_{l=0}^{\infty} \sum_{m=-l}^{l} a_{lm} Y_{lm}(\theta, \phi).
    \label{eq:2}
\end{equation}
Using Eqn. (\ref{eq:1}) and Eqn. (\ref{eq:2}) we write the ratio of  \( \kappa_\perp \)(\( \theta = \frac{\pi}{2} \)) to \( \kappa_{\parallel} \)(\( \theta = 0 \)) as :

\begin{equation}
\label{eq:3}
f\left( \frac{\pi}{2}, \phi \right) = \frac{\kappa_\perp}{\kappa_\parallel} = 
\frac{
\displaystyle\sum_{l=0}^{\infty} \sum_{m=-l}^{l} a_{lm} \, Y_{lm}\left(\frac{\pi}{2}, \phi\right)
}{
\displaystyle\sum_{l=0}^{\infty} \sum_{m=-l}^{l} a_{lm} \, Y_{lm}(0, \phi)
}
\end{equation}
which can be simplified as 
\begin{equation}
\label{eq:4}
\kappa_\perp = f\left( \frac{\pi}{2}, \phi \right) \cdot \kappa_\parallel \ \ ,
\end{equation} 
This relation is satisfied by all types of opacity, but it is preferred to use parallel and perpendicular mean opacities (Rosseland mean/Planck mean), depending on optically thin/thick media, to encode all kinds of micro-physical interactions and all other properties of anisotropic media. The expansion of spherical harmonics can thus be constrained by using pre-computed opacity tables for varying temperature, magnetic field, and density profiles, depending on the choice of stellar profile. Choosing a particular symmetry and complexity of angular system we can truncate the infinite series to a finite number of terms. For practical computation, we can solve the system maintaining physical admissibility. Choosing a particular range of temperature profiles to analyze will also give an estimate of the heat flux and $\alpha$ in numerical analysis. As all kinds of interactions in anisotropic media are covered in mean opacities, we do not need to add additional magnetic pressure while analyzing the collapse conditions. We will consider here a highly magnetized (source of anisotropies) and radiation-dominated regime, from which we can write the relation between pressure and opacity components as 
\begin{align}
P_{\perp} &= C \cdot \kappa_{\perp} 
\label{eq:Pperp} \\
P_{\parallel} &= C \cdot \kappa_{\parallel} 
\label{eq:Pparallel}
\end{align}
where $C$ is the proportionality constant with the units setting equivalence to the relation because in radiation-dominated regimes, opacity is directly proportional to the pressure components \cite{rybicki1979radiative}. Now using Eqns. (\ref{eq:4} , \ref{eq:Pperp} , and \ref{eq:Pparallel}), we can write
\begin{equation}\label{eq:7}
 \Delta = P_t - P_r =  P_{\perp}-P_{\parallel} = C.\kappa_\parallel(f\left( \frac{\pi}{2}, \phi \right) - 1)
\end{equation}
\section{Opacity-Controlled Collapse and Final Outcomes}
The detailed physical conditions and the corresponding anisotropy-dependent collapse outcomes have been comprehensively summarized in Table~\ref{tab:anisotropy_conditions}. Here, we proceed to update the critical inequalities using the new microphysically derived expression for the anisotropic stress. Specifically, the anisotropic stress is now expressed as:
\begin{equation}
    \Delta = C \, \kappa_{\parallel} \left[ f \left( \frac{\pi}{2}, \phi \right) - 1 \right]
\end{equation}
where \( f \left( \frac{\pi}{2}, \phi \right) \) is the dimensionless geometric factor encoding the angular variation of opacity, as extracted from stellar opacity data.

Substituting this expression into the critical inequalities yields the following directional threshold conditions, where the geometric factor is now explicitly isolated. These inequalities are valid under their corresponding dynamical collapse conditions.

\begin{widetext}

\begin{table}[H] 
\centering
\renewcommand{\arraystretch}{1.6}
\begin{tabular}{|c|p{6.5cm}|p{9cm}|}
\hline
\textbf{Outcome} & \textbf{Physical Conditions} & \textbf{Directional Threshold Condition} \\
\hline

\textbf{Black Hole} & 
\begin{itemize}
    \item $U < 0$ (collapse proceeds)
    \item $D_t U < 0$ (collapse accelerates)
\end{itemize} 
& 
$\displaystyle 
f\!\left( \frac{\pi}{2}, \phi \right) < 1 + \frac{1}{C \kappa_{\parallel}} 
\frac{R}{2 E^2}
\left[
\begin{aligned}
    & (\mu + P_r + 2 \epsilon) 
      \left( \frac{m + 4 \pi (P_r + \epsilon) R^3}{R^2} \right) \\
    & + E^2 \left( D_R (P_r + \epsilon) + \frac{2 \epsilon}{R} \right) \\
    & + E \left( 5B(q + \epsilon)\frac{U}{R} 
      + B D_t q + D_t \epsilon \right)
\end{aligned}
\right]$
\\
\hline

\textbf{Naked Singularity} & 
\begin{itemize}
    \item $U < 0$ (collapse continues)
    \item $D_t U > 0$ (decelerating collapse)
    \item Trapping delayed beyond singularity
\end{itemize} 
& 
$\displaystyle 
f\!\left( \frac{\pi}{2}, \phi \right) > 1 + \frac{1}{C \kappa_{\parallel}} 
\frac{R}{2 E^2}
\left[
\begin{aligned}
    & (\mu + P_r + 2 \epsilon) 
      \left( \frac{m + 4 \pi (P_r + \epsilon) R^3}{R^2} \right) \\
    & + E^2 \left( D_R (P_r + \epsilon) + \frac{2 \epsilon}{R} \right) \\
    & + E \left( 5B(q + \epsilon)\frac{U}{R} 
      + B D_t q + D_t \epsilon \right)
\end{aligned}
\right]$
\\
\hline

\textbf{Bounce} & 
\begin{itemize}
    \item $U = 0$ (collapse halts)
    \item $D_t U > 0$ (outward acceleration begins)
\end{itemize} 
& 
$\displaystyle 
f\!\left( \frac{\pi}{2}, \phi \right) > 1 + \frac{1}{C \kappa_{\parallel}} 
\frac{R}{2 E^2}
\left[
\begin{aligned}
    & (\mu + P_r + 2 \epsilon) 
      \left( \frac{m + 4 \pi (P_r + \epsilon) R^3}{R^2} \right) \\
    & + E^2 \left( D_R (P_r + \epsilon) + \frac{2 \epsilon}{R} \right) \\
    & + E \left( B D_t q + D_t \epsilon \right)
\end{aligned}
\right]$
\\
\hline
\end{tabular}

\caption{Physical conditions and directional threshold criteria determining the final outcomes of anisotropic-dissipative gravitational collapse. The function $f(\theta, \phi)$ represents the directional geometric factor, while $\kappa_{\parallel}$ denotes the longitudinal thermal conductivity coefficient.}
\label{tab:directional_conditions}
\end{table}

\end{widetext}

\section{Results and conclusions}
\label{sec:discussion}
We now have a collapse threshold as a function of directionally dependent micro-physically rich radiative opacity encoded in $f \left( \frac{\pi}{2}, \phi \right)$ as defined in equation (\ref{eq:3}). In this study we have demonstrated that the dynamics of gravitational collapse can be significantly influenced by the microphysics of the stellar interior, particularly through the role of directional opacity. We have shown that opacity-induced anisotropy is capable of slowing, halting, or even reversing collapse in specific directions. Importantly, the anisotropies and dissipation in our framework emerge from physically realistic opacity structures rather than being imposed artificially. This connects the fate of collapse not only to the spacetime geometry, but also to the thermodynamic and radiative properties of the star. While our model does not explicitly track the formation of trapped surfaces via null geodesic integration or apparent horizon conditions, it clearly establishes the dynamical conditions under which such phenomena could be delayed. To support this approach, we refer the reader to {Appendix~\ref{appendix:trapped}}, where established results from the literature justify how dissipation and high-opacity environments can delay or prevent trapped surface formation. These works reinforce the physical plausibility of the outcomes derived in our opacity-controlled collapse model. Several directions emerge from this work:

\begin{itemize}
  \item Numerical implementation using precomputed opacity tables, which are inherently time-independent, can allow detailed simulations of collapse outcomes in different astrophysical environments.
  \item Stellar profile mapping based on observational data or simulation outputs could be used to determine regions where opacity anisotropies might result in directional deceleration or even bounce.
  \item Extension to rotating or axially symmetric configurations could reveal more complex collapse behavior under anisotropic pressure and magnetic field-aligned opacity structures.
  \end{itemize}

The opacity-governed approach presented here offers a novel and physically motivated lens through which to study gravitational collapse. It bridges thermodynamics, radiation transport, and general relativity, offering a broader picture of how stars may meet their end — not uniformly, but directionally directed by the very light they struggle to emit.
\section{Acknowledgments}

I would like to express my sincere gratitude to Oem Trivedi for his invaluable guidance, mentorship, and encouragement throughout this research journey and 
for endorsing me for gr-qc at arxiv. I am also sincerely thankful to Meet Vyas, Kharanshu Solanki, Professor Pankaj Joshi, and Prfoessor Avi Loeb for their insightful discussions and constructive feedback, which helped me to refine the presentation and broaden the context of this manuscript.

\newpage
\begin{widetext}
    
\appendix
\renewcommand{\thesection}{\Alph{section}}
\section{Appendix A: Justification for Trapped Surface Delay Without Geodesic Analysis}
\label{appendix:trapped}

While we do not explicitly solve for null geodesics or apparent horizon formation, the dynamical impact of dissipation and anisotropy on collapse has been previously studied and established. Joshi and Goswami \cite{JoshiGoswami2007} have discussed in detail about the formation and delay of trapped surfaces, including radial heat flux and dissipation can act as a bulk mechanism to extract mass from the interior and delay or prevent trapped surface formation. Their discussion involves realistic dissipative matter and matching a Vaidya exterior spacetime, consistent with the present model. In such scenarios, outgoing energy flux weakens gravitational trapping, allowing locally naked singularities to emerge.

In particular, Banerjee, Chatterjee, and Dadhich~\cite{Banerjee2002} demonstrated that a spherically symmetric dissipative configuration with non-vanishing heat flux can undergo continual collapse without the formation of an event horizon. Their analysis showed that the rate of mass loss is balanced by the fall of boundary radius and hence if we only track the collapsing velocity and acceleration we can check whether the end state is a naked singularity or a blackhole.  This idea was later extended by Banerjee and Chatterjee~\cite{Banerjee2005} to higher dimensions, where the same mechanism persists and the inclusion of additional dimensions further delays the onset of trapping. Both studies emphasize the crucial roles of pressure anisotropy and heat flux in governing the final fate of collapse .

Furthermore, Herrera \cite{herrera2009} has demonstrated that in high-opacity, high-temperature environments relevant to stellar interiors, the critical factor from Eqn. (\ref{eq:alpha}) can approach unity. As \( \alpha \rightarrow 1 \), the effective inertial mass density vanishes or becomes negative, leading to significant collapse deceleration or even bounce. This mechanism can halt collapse or delay the formation of trapped surfaces.

Given that pressure anisotropy in our model is explicitly sourced by directional opacity and that the critical factor-corrected inertia is built into our dynamical framework, these results provide strong physical justification for the trapping delay and directional visibility conditions we describe.

Therefore, the present analysis based on the dynamical acceleration of the collapse is a self-contained criterion for determining the end state of anisotropic-dissipative gravitational collapse. By isolating the anisotropic stress term and embedding the angularly dependent opacity into the critical inequalities as derived in table \ref{tab:anisotropy_conditions}, the necessary and sufficient conditions for horizon formation, bounce, or naked singularity emerge directly from the local dynamics. This approach eliminates the need for an explicit null-geodesic integration or horizon-tracking analysis, since the sign of the acceleration and boundary velocity already encapsulates the physical outcome.

\end{widetext}

\bibliographystyle{apsrev4-2}
\nocite{*}
\bibliography{references}

\end{document}